\documentclass[twocolumn]{aastex62}

\usepackage{natbib}
\usepackage{threeparttable}
\usepackage{booktabs}
\usepackage{amsmath}
\usepackage{enumitem}

\received{}
\revised{}
\accepted{}

\shorttitle{Evidence for a threshold icm density}
\shortauthors{Roberts et al.}

\begin{document}

\title{Quenching low-mass satellite galaxies: evidence for a threshold ICM density}

\correspondingauthor{Ian D. Roberts}
\email{roberid@mcmaster.ca}

\author[0000-0002-0692-0911]{Ian D. Roberts}
\affil{Department of Physics \& Astronomy, McMaster University \\
Hamilton ON L8S 4M1, Canada}

\author[0000-0003-4722-5744]{Laura C. Parker}
\affil{Department of Physics \& Astronomy, McMaster University \\
Hamilton ON L8S 4M1, Canada}

\author[0000-0003-1845-0934]{Toby Brown}
\affil{Department of Physics \& Astronomy, McMaster University \\
Hamilton ON L8S 4M1, Canada}

\author{Gandhali D. Joshi}
\affil{Max Planck Institute for Astronomy \\
D-69117 Heidelberg, Germany}

\author{Julie Hlavacek-Larrondo}
\affil{D{\'e}partement de Physique, Universit{\'e} de Montr{\'e}al \\
Montr{\'e}al QC H3C 3J7, Canada}

\author[0000-0001-8745-0263]{James Wadsley}
\affil{Department of Physics \& Astronomy, McMaster University \\
Hamilton ON L8S 4M1, Canada}



\begin{abstract}
We compile a sample of SDSS galaxy clusters with high-quality \textit{Chandra} X-ray data to directly study the influence of the dense intra-cluster medium (ICM) on the quenching of satellite galaxies.  We study the quenched fractions of satellite galaxies as a function of ICM density for low- ($10^9 \lesssim M_\star \lesssim 10^{10}\,\mathrm{M_\odot}$), intermediate- ($10^{10} \lesssim M_\star \lesssim 10^{10.5}\,\mathrm{M_\odot}$), and high-mass ($M_\star \gtrsim 10^{10.5}\,\mathrm{M_\odot}$) satellite galaxies with $>\!3000$ satellite galaxies across 24 low-redshift ($z < 0.1$) clusters.  For low-mass galaxies we find evidence for a broken powerlaw trend between satellite quenched fraction and local ICM density.  The quenched fraction increases modestly at ICM densities below a threshold before increasing sharply beyond this threshold toward the cluster center.  We show that this increase in quenched fraction at high ICM density is well matched by a simple, analytic model of ram pressure stripping.  These results are consistent with a picture where low-mass cluster galaxies experience an initial, slow-quenching mode driven by steady gas depletion, followed by rapid quenching associated with ram pressure of cold-gas stripping near (one quarter of the virial radius, on average) the cluster center.
\end{abstract}

\keywords{Galaxies: clusters: general, Galaxies: clusters: intracluster medium, Galaxies: evolution, Galaxies: star formation, X-rays: galaxies: clusters}


\section{Introduction} \label{sec:intro}

It is now firmly established that local environment plays a pivotal role in dictating the properties of galaxy populations.  Galaxies located in the underdense field tend to be blue in colour, with young stellar populations, disc (late-type) morphologies, and high star formation rates (SFRs).  In contrast, dense environments such as galaxy clusters, embedded within massive ($\gtrsim 10^{14}\,\mathrm{M_\odot}$) dark matter halos, host galaxy populations which are on average red, with bulge-dominated (early-type) morphologies, old stellar populations, and little ongoing star formation.  The first clear evidence for such a paradigm was presented in early seminal works \citep[e.g.][]{oemler1974, butcher1978, davis1976, dressler1980, postman1984}, and has since been cemented by more recent studies using large, detailed spectroscopic surveys of galaxies across a variety of environments \citep[e.g.][]{blanton2009, kimm2009, peng2010, wetzel2012, wilman2012}.  Even within individual clusters, galaxy properties are a strong function of environment.  Galaxies that inhabit the dense cluster interior are preferentially red, early-type, and quiescent relative to galaxies at large cluster-centric radius \citep[e.g.][]{postman2005, blanton2007, prescott2011, rasmussen2012, fasano2015, haines2015}.  Ultimately, if we are to understand galaxy evolution we must understand which physical mechanisms drive these observed trends by quenching star formation and transforming morphology as a function of galaxy environment.
\par
In addition to environmental quenching occuring in a galaxy's present day cluster, there is increasing evidence that a large fraction of galaxies may have their star formation quenched in smaller groups prior to cluster infall.  It has been estimated that nearly half of present day cluster galaxies may have infallen as a part of a smaller groups \citep[e.g.][]{mcgee2009}, and quenched fractions at the cluster virial radius tend to be enhanced relative to the field -- indicating that star formation is influenced environmentally prior to infall \citep{vonderlinden2010, haines2015, roberts2017}.  This ``pre-processing'' of galaxy properties is a natural consequence of a hierachical growth of dense structures in the Universe.
\par
Many physical mechanisms have been proposed which, in principle, are capable of quenching star formation in dense environments.  These mechanisms can be broadly divided into two classes: hydrodynamic interactions between galaxies and the intracluster medium (ICM); and dynamical interactions between member galaxies, or between galaxies and the cluster halo potential.  Examples of hydrodynamic mechanisms include: `starvation' \citep[e.g.][]{larson1980, balogh2000, peng2015}, where the high virial temperature of the cluster ($\gtrsim 10^7\,\mathrm{K}$) prevents cold-flow accretion of gas onto the disc of satellite galaxies; and `ram pressure stripping' \citep[e.g.][]{gunn1972, quilis2000}, where a galaxy passing through the dense ICM will feel a ram pressure `wind' which is strong enough to directly strip cold-gas from the galactic disc.  Dynamical interactions thought to be relevant include: galaxy mergers \citep[e.g.][]{mihos1994a, mihos1994b}, where the final end products in dense environments tend to be quiescent galaxies with early-type morphologies; `harassment' \citep[e.g.][]{moore1996}, where repeated impulsive interactions between galaxies can induce strong starbursts, thereby quickly exhausting cold-gas reserves; and gravitational tidal forces \citep[e.g.][]{mayer2006, chung2007}, which can directly strip gas from a galaxy, or in the less extreme case, transport gas to less bound orbits where it will be more susceptible to hydrodynamic effects such as ram pressure.  While it is understood that all of the aforementioned processes should be affecting galaxies in dense environments, it is the balance between these mechanisms, and the dependence of this balance on environment that fuels substantial debate.
\par
Recently, starvation or ram pressure stripping (or a combination of the two) have been favoured as the primary quenching mechanism in galaxy groups and clusters \citep{muzzin2014, peng2015, fillingham2015, wetzel2015, brown2017, foltz2018}.  Substantial effort has been devoted to determine how to distinguish between these two quenching pathways observationally.  A common technique is to constrain the timescale over which quenching occurs, which is expected to be relatively long ($\gtrsim 3-4\,\mathrm{Gyr}$) for starvation but short ($\lesssim 1\,\mathrm{Gyr}$) for ram pressure stripping.  It is important to note, however, that ram pressure stripping of cold-gas will not be immediately efficient upon cluster infall and a delay time is likely necessary for the galaxy to reach the dense interior ICM before quenching begins.  Therefore the total quenching timescale for ram pressure (delay + quenching) should be on the order of the cluster dynamical time.  Starvation should begin to act immediately after infall as the galaxy encounters the hot, virialized halo, however quenching by starvation will still produce an excess of quenched galaxies in the cluster interior (relative to the outskirts) as the time-since-infall for these central galaxies will be relatively long.
\par
The effects of ram pressure stripping can, in some cases, be studied directly by observing cluster galaxies with extended \textsc{Hi} distributions \citep{kenney2004, chung2007, chung2009, kenney2015}, \textsc{Hi} deficient discs (post-stripping, \citealt{kenney1989, boselli2006, jaffe2016}), and by observing ``jellyfish galaxies'' with extended, stripped ``tentacles'' of gas and stars \citep{poggianti2017, jaffe2018}.  Ram pressure also lends itself well to analytic modelling through the simple balance between the restoring potential of a galactic disc and the strength of ram pressure given by $\rho_\mathrm{ICM}\,v_\mathrm{galaxy}^2$ \citep{gunn1972}.  Such an approach has been used to constrain the regions in cluster phase space where stripping should be efficient, finding that the ``stripping'' regions tend to be populated by galaxies which are \textsc{Hi}-deficient and show morphological signs of ongoing stripping \citep{jaffe2015, jaffe2016, jaffe2018}.
\par
In this paper we compile a statistical sample ($>3000$) of satellite galaxies in 24  low redshift ($z < 0.1$) clusters observed with the \textit{Chandra} X-ray observatory to directly study the connection between galaxy quenching and ICM density.  We can then estimate the physical ICM density (based on density profiles for each cluster) around each satellite galaxy and constrain the impact of ram pressure stripping using a simple analytic model.  This represents the first systematic study of environmental quenching directly as a function ICM density for a large, statistical sample of galaxies across many clusters.
\par
The paper is organized as follows: in section~\ref{sec:data} we describe the sample of cluster satellite galaxies, as well as the optical and X-ray data for the host clusters; in section~\ref{sec:quench_dens} we present the dependence of satellite quenched fraction on ICM density; in section~\ref{sec:rp_model} we describe an analytic ram pressure stripping model we construct and make comparisons to the observed trends; in section~\ref{sec:slow-then-rapid} we constrain quenching timescales assuming a ``slow-then-rapid'' framework for satellite quenching; and finally, in sections~\ref{sec:discussion} and \ref{sec:summary} we discuss and summarize our results.
\par
This paper assumes a flat $\mathrm{\Lambda}$ cold dark matter cosmology with $\Omega_\mathrm{M} = 0.3$, $\Omega_\mathrm{\Lambda} = 0.7$, and $H_0 = 70\,\mathrm{km}\,\mathrm{s^{-1}}\,\mathrm{Mpc^{-1}}$.  Throughout this paper we will use lowercase $r$ to represent galactocentric radii (in cylindrical coordinates), and uppercase $R$ to represent cluster-centric radii.

\section{DATA}
\label{sec:data}

\subsection{Cluster sample}
\label{sec:clusters}

\begin{table*}
  \scriptsize
  \caption{Galaxy cluster sample}
  \label{tab:sample}
  \centering
  \begin{threeparttable}
  \begin{tabular}{lcccccccl}
    \toprule
    Name & Yang ID$^a$ & $z_\mathrm{cluster}$$^b$ & $M_\mathrm{halo}$$^c$ & $R_{500}$ & $L_{X,\mathrm{bol}}$$^d$ & $N_\mathrm{gal}/N_\mathrm{Yang}$$^e$ & Exp. time & \textit{Chandra} ObsID \\[1.5pt]
    & & & $(10^{14}\,\mathrm{M_\odot})$ & (kpc) & $(10^{44}\,\mathrm{erg}\,\mathrm{s^{-1}})$& & (ks) & \\
    \midrule
    Coma & 1 & 0.024 & 10.8 & 1150 & 7.1$^*$ & 517/652 & 479 & 13993,13994,13995 \\
    &&&&&&&& 13996,14406,14410 \\
    &&&&&&&& 14411,14415 \\[1.5pt]
    Abell 2147 & 2 & 0.036 & 9.6 & 1078 & 0.7 & 329/383 & 18 & 3211 \\[1.5pt]
    Abell 1367 & 3 & 0.022 & 6.2 & 955 & 1.2$^*$ & 223/352 & 402 & 514,17199,17200 \\
    &&&&&&&& 17201,17589,17590 \\
    &&&&&&&& 17591,17592 \\
    &&&&&&&& 18704,18705,18755 \\[1.5pt]
    Abell 2199 & 5 & 0.030 & 5.8 & 920 & 0.8 & 221/302 & 158 & 497,498,10748 \\
    &&&&&&&& 10803,10804,10805 \\[1.5pt]
    Abell 85 & 11 & 0.056 & 8.4 & 1030 & 5.3 & 152/178 & 198 & 904,15173,15174 \\
    &&&&&&&& 16263,16264 \\[1.5pt]
    Abell 2063 & 18 & 0.035 & 3.8 & 799 & 4.3 & 132/156 & 50 & 4187,5795,6262 \\
    &&&&&&&& 6263 \\[1.5pt]
    Abell 2670 & 19 & 0.076 & 12.6 & 1115 & 2.3$^*$ & 107/154 & 40 & 4959 \\[1.5pt]
    Abell 2029 & 20 & 0.077 & 9.4 & 1041 & 15.6 & 114/154 & 128 & 891,4977,6101 \\[1.5pt]
    Abell 2065 & 21 & 0.072 & 8.6 & 1013 & 3.0 & 104/154 & 55 & 3182,7689 \\[1.5pt]
    Abell 2142 & 22 & 0.090 & 17.6 & 1259 & 75.9 & 149/153 & 205 & 5005,15186,16564 \\
    &&&&&&&& 16565 \\[1.5pt]
    MKW 8 & 23 & 0.027 & 1.2 & 697 & 0.9 & 132/150 & 104 & 4942,18266,18850 \\[1.5pt]
    Abell 2107 & 24 & 0.041 & 3.8 & 787 & 3.0 & 119/149 & 36 & 4960 \\[1.5pt]
    Abell 2052 & 25 & 0.035 & 4.0 & 815 & 5.1 & 103/141 & 654 & 890,5807,10477 \\
    &&&&&&&& 10478,10479,10480 \\
    &&&&&&&& 10879,10914,10915 \\
    &&&&&&&& 10916,10917 \\[1.5pt]
    Abell 2255 & 27 & 0.082 & 13.6 & 1167 & 3.7 & 107/130 & 44 & 894,7690 \\[1.5pt]
    Abell 2061 & 29 & 0.078 & 5.1 & 1072 & 2.5$^*$ & 115/129 & 55 & 4965 \\[1.5pt]
    Abell 1795 & 32 & 0.063 & 5.8 & 895 & 7.8 & 103/125 & 105 & 493,494,10432,17228 \\[1.5pt]
    ZwCl 1215 & 45 & 0.077 & 7.3 & 954 & 3.5 & 93/103 & 12 & 4184 \\[1.5pt]
    Abell 1991 & 52 & 0.058 & 5.2 & 862 & 0.5 & 96/99 & 38 & 3193 \\[1.5pt]
    MKW 3S & 57 & 0.045 & 3.1 & 743 & 1.1 & 77/95 & 57 & 900 \\[1.5pt]
    MKW 4 & 62 & 0.021 & 1.7 & 625 & 0.03 & 58/88 & 30 & 3234 \\[1.5pt]
    Abell 1775 & 71 & 0.075 & 5.7 & 880 & 2.8$^*$ & 70/82 & 99 & 12891,13510 \\[1.5pt]
    Abell 1650 & 86 & 0.084 & 7.3 & 947 & 4.4 & 56/72 & 251 & 4178,5822,5823 \\
    &&&&&&&& 6356,6357,6358 \\
    &&&&&&&& 7242,7691 \\[1.5pt]
    AWM 4 & 145 & 0.032 & 0.7 & 564 & 0.4$^*$ & 44/55 & 74 & 9423 \\
    Abell 2244 & 191 & 0.098 & 4.7 & 805 & 5.5 & 44/47 & 65 & 4179,7693 \\
    &&&&&&&& 13192,13193 \\[1.5pt]
    NGC 4325 & 611 & 0.026 & 0.3 & 336 & 0.2$^*$ & 18/24 & 30 & 3232 \\[1.5pt]
    \bottomrule
  \end{tabular}
  \begin{tablenotes}
    \footnotesize
    \item \textsc{Notes.} $^a$Group ID from sample III in the Yang catalogue \cite{yang2007}; $^b$Cluster redshift from Yang catalogue; $^c$Halo mass from Yang catalogue determined using ranking of cluster stellar mass; $^d$Bolometric X-ray luminosities taken from the ACCEPT catalogue \cite{cavagnolo2009} when available, and otherwise taken from the ROSAT all-sky survey (marked by $^*$, \citealt{wang2014}) ; $^e$Number of cluster members in the final sample after matching with SFRs and B+D decompositions, compared to the number of cluster members identified in the Yang catalogue.
  \end{tablenotes}
  \end{threeparttable}
  
\end{table*}

\begin{figure*}
	\centering
	\includegraphics[width=\textwidth]{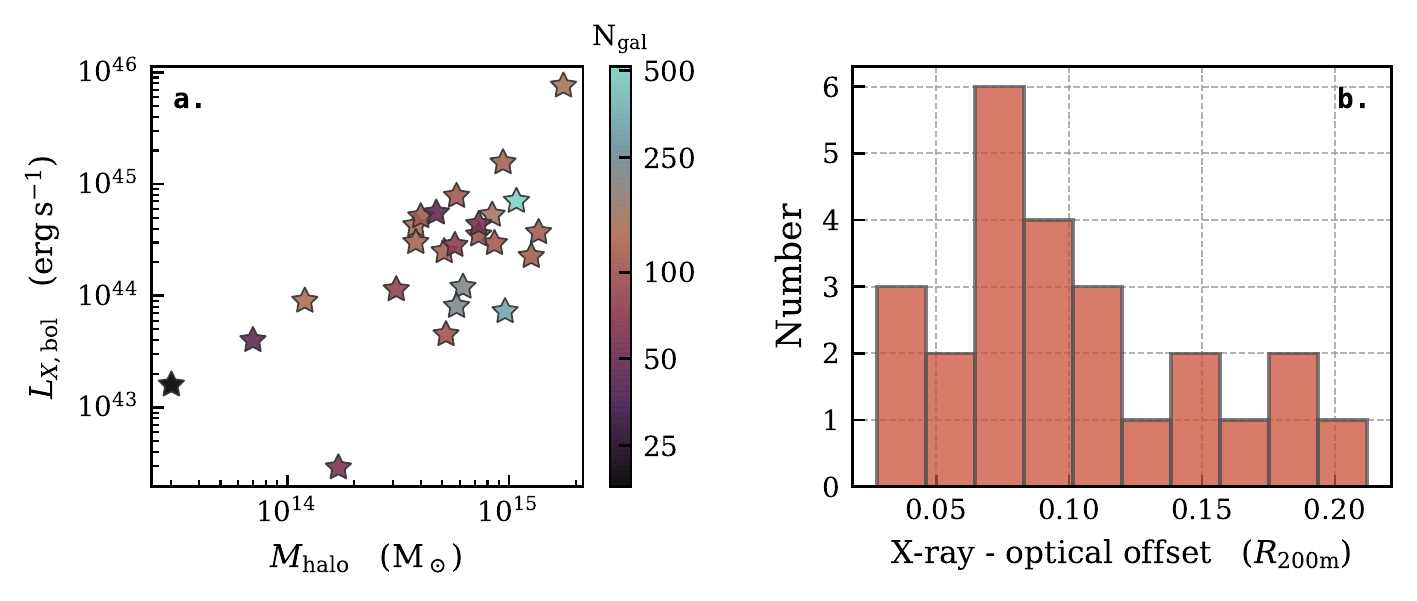}
	\caption{\textit{Left:}  Literature bolometric X-ray luminosity (from the ACCEPT cluster sample when available: \citealt{cavagnolo2009}, otherwise from the SDSS-RASS sample: \citealt{wang2014}) versus halo mass \citep{yang2007} for the 24 clusters in the sample.  Markers are coloured according to the number of galaxies identified in each cluster.  \textit{Right:}  Projected offset between the X-ray peak and optical luminosity-weighted centre for each cluster in the sample.}
	\label{fig:opt_xray_offset}
\end{figure*}

We construct a sample of low redshift galaxy clusters with high quality, archival X-ray observations, starting with all clusters in the \citet{yang2005, yang2007} SDSS DR7 catalogue at $z < 0.1$ with ten or more member galaxies.  We then query the \textit{Chandra} data archive at the positions of the luminosity-weighted centres of the Yang clusters in this initial sample.  We require that any matches have at least 25000 X-ray counts above the background (after combining all \textit{Chandra} archival observations for a given cluster), which results in 24 clusters that both are in the \citeauthor{yang2007} catalogue and are observed with \textit{Chandra} to sufficient depth.  Table~\ref{tab:sample} lists the clusters in this sample, Fig~\ref{fig:opt_xray_offset}a shows the $L_X - M_\mathrm{halo}$ relation for these clusters, and Fig.~\ref{fig:opt_xray_offset}b shows the projected offset between the position of the X-ray peak and the luminosity-weighted cluster centre, showing that typical offsets are only a small fraction of the virial radius (see equation~\ref{eq:Rvir}).  X-ray centres
are calculated as the position of the brightest pixel in
the X-ray image after smoothing using a Gaussian kernel
with a bandwidth of $40\,\mathrm{kpc}$ \citep{nurgaliev2013, roberts2018} and luminosity-weighted centres are computed using the positions of member galaxies from the \citeauthor{yang2007} catalogue.  Bolometric X-ray luminosities are archival and are taken from the ACCEPT cluster sample \citep{cavagnolo2009} when available, and otherwise from the Rosat All Sky Survey \citep{wang2014}.  Cluster halo masses are taken from the \citeauthor{yang2007} catalogue which are computed using abundance matching based on the ranking of the characteristic group stellar mass, given by equation 13 in \citet{yang2007}.  We normalize all cluster-centric radii by $R_{500}$ (the radius at which the interior density is 500 times the critical density of the Universe) or by $R_{200m}$, which are computed as
\begin{equation}
  R_{500} = R_{200m} / 2.7
\end{equation}
\noindent
where,
\begin{equation}\label{eq:Rvir}
  R_{200m} = 1.61\,\mathrm{Mpc}\,\left(\frac{M_\mathrm{halo}}{10^{14}\,\mathrm{M_\odot}}\right)^{1/3}\,(1 + z_\mathrm{cluster})^{-1}
\end{equation}
\noindent
is the radius enclosing an average density equal to 200 time the critical mass density of the Universe \citep{yang2007, tinker2008, wang2014}.

\subsubsection{ICM density profiles}
\label{sec:dens_profs}

\begin{figure}
	\centering
	\includegraphics[width=\columnwidth]{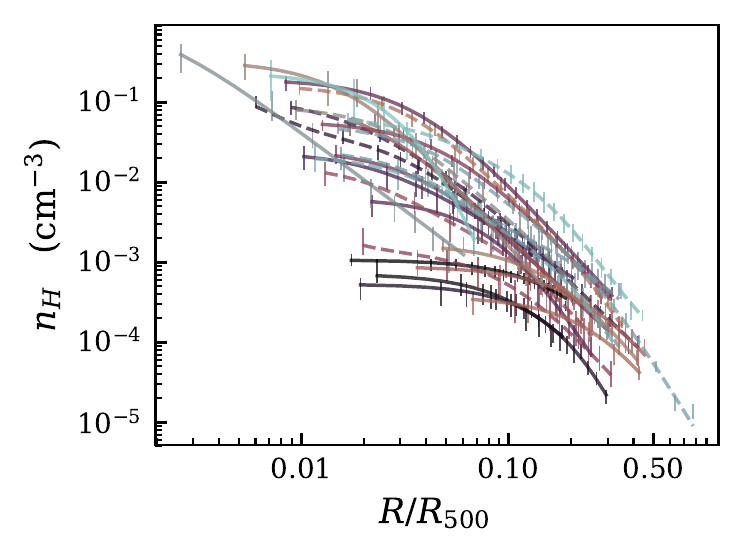}
	\caption{Best-fit ICM density profile, assuming a single or double beta model (whichever gives the lowest AIC), for each cluster in the sample.  Clusters which are better fit by a double-beta model are marked as dashed lines.  Error bars correspond to 1$\sigma$ statistical uncertainties on the measured densities.}
	\label{fig:dens_profiles}
\end{figure}

The \textit{Chandra} observations were downloaded, reprocessed, cleaned, and calibrated using \textsc{ciao} (\textsc{ciao} version 4.9, \textsc{caldb} version 4.7.7).  We apply charge transfer inefficiency and time-dependent gain corrections and filter for background flares using the \textsc{ciao} script \textsc{lc\_clean} with a $3\sigma$ threshold. Point sources are identified and masked using the \textsc{wavdetect} script. Backgrounds were estimated using the blanksky event file output from the \textsc{ciao} script \textsc{blanksky} which is normalized to the ratio of observed-to-blanksky background counts in the $9 - 12\,\mathrm{keV}$ band.  To determine the ICM density as a function of radius we extract X-ray spectra in radial annuli from the source and the background data sets using the \textsc{ciao} script \textsc{specextract} in the $0.5 - 7\,\mathrm{keV}$ energy band.  Annuli are centered on the X-ray peak and extend out to the edge of the chip coverage.  Weighted response files and redistribution matrices were generated with a count-weighted map across the extent of the extraction regions.  We then fit the X-ray spectra in radial annuli for each cluster in the sample.  We set a minimum of 5000 counts (after background subtraction) per annulus, which is based on the merged dataset for clusters with multiple observations (see table~\ref{tab:sample}), and then group spectra to have 25 counts per energy channel (ie.\ $\mathrm{S/N = 5}$).  We only consider clusters with at least 5 radial annuli (ie.\ 25000 counts), and set an upper limit of 20 annuli per cluster -- this means that clusters with deep \textit{Chandra} observations will have far more than 5000 counts per annulus.  To gain insight into the physical densities in each annulus, we deproject the spectra using the code \textsc{dsdeproj} \citep{sanders2007, russell2008}, which is a model-independent deprojection method assuming only spherical symmetry.  We then fit the deprojected spectrum in each annulus with an absorbed single-temperature \textsc{apec} model\footnote{http://www.atomdb.org/}, with temperature, abundance, and normalization as free parameters.  For clusters with multiple \textit{Chandra} observations we extract a spectrum for each dataset and then simultaneously fit all spectra in \textsc{sherpa}\footnote{http://cxc.harvard.edu/sherpa/}.  The redshift is fixed at the cluster redshift from the Yang catalogue. The Galactic hydrogen column density is estimated from the spectrum extracted over the entire cluster region, which gives sufficient counts to obtain a good constraint.  We then assume that the Galactic hydrogen column density ($N_H$) is constant across the cluster and fix $N_H$ at this fitted value for the spectral fits within each radial annulus.  The fitted hydrogen column densities agree well with the observed values from \citet{kalberla2005}.  The spectral normalization is then converted into a hydrogen number density using the following relation,
\begin{equation}
  n_H = \sqrt{\frac{4 \pi D_A^2 (1 + z)^2 \eta \cdot 10^{14}}{1.2 V}}
\end{equation}
\noindent
where $D_A$ is the angular diameter distance to the cluster at redshift, $z$, $\eta$ is the spectral normalization, $V$ is the volume corresponding to a given annulus, and we have assumed $n_e = 1.2\,n_H$.  The deprojected density profile for each cluster is then fit with both a single (equation~\ref{eq:beta}) and a double (equation~\ref{eq:beta2}) beta model,
\begin{equation} \label{eq:beta}
  n_H = n_{H,0} \left[1 + \left(\frac{R}{R_c}\right)^2\right]^{-\frac{3}{2}\beta}
\end{equation}
\begin{equation}\label{eq:beta2}
\begin{split}
  n_H &= n_{H,01} \left[1 + \left(\frac{R}{R_{c1}}\right)^2\right]^{-\frac{3}{2}\beta_1}\\
  &+n_{H,02} \left[1 + \left(\frac{R}{R_{c2}}\right)^2\right]^{-\frac{3}{2}\beta_2}
\end{split}
\end{equation}
\noindent
with the central density, $n_{H,0}$, core radii, $R_c$, and beta indices, $\beta$, as free parameters.  Given the single and double beta fit for each cluster, we use the fit which gives the lowest Akaike information criterion\footnote{The Akaike information criterion is a model selection tool used to quantify the information lost when describing a set of data with a particular model.  When comparing two models, the model with the lowest Akaike information criterion value is preferred.  The Akaike information criterion also includes a penalty for increasing the number of fits parameters, thereby accounting for obtaining ``better'' fits by increasing model complexity.} (AIC; \citealt{akaike1974}) value.  Allowing clusters to be parameterized by a double-beta model when necessary allows for the more accurate modelling of systems with strong cool-cores, however we note that using a single-beta model for all clusters in the sample does not change the conclusions of this paper. The data prefer a double-beta model for 8 of the clusters in the sample, with the remaining 16 preferring a single-beta fit.  The ICM density profile fits are done using the MCMC code \textsc{emcee}\footnote{http://dfm.io/emcee/current/} \citep{foreman-mackey2013} assuming a flat prior; best-fit parameters are taken to be the median values from each chain after burn-in. Fig.~\ref{fig:dens_profiles} shows the best-fit profiles for each cluster as well as 1$\sigma$ error bars corresponding to the measured densities.  These fits allow us to determine the local, azimuthally averaged, ICM density for each satellite galaxy at a given cluster-centric radius (extrapolating the fits to larger cluster-centric radii when necessary).

\subsection{Galaxy sample}
\label{sec:galaxies}

We compile a sample of cluster satellite galaxies beginning with the member galaxies for the clusters in Table~\ref{tab:sample} from the \citeauthor{yang2007} DR7 catalogue (we remove the most-massive galaxy from each cluster).  Galaxies are then matched to star formation rates (SFRs) from the Max Planck Institut f\"{u}r Astrophysik and Johns Hopkins University (MPA-JHU) collaboration\footnote{https://www.mpa-garching.mpg.de/SDSS/DR7} \citep{brinchmann2004}, with the updated prescriptions from \citet{salim2007}.  To determine specific star formation rates ($\mathrm{sSFR} = \mathrm{SFR} / M_\star$) we use stellar masses from \citet{mendel2014} derived via fits to galaxy broadband spectral energy distributions (SEDs).  \citeauthor{mendel2014} also derive bulge and disc stellar masses using the bulge+disc (B+D) decompositions from \citet{simard2011} (assuming an exponential disc and a De Vaucouleurs bulge), which we use in our ram pressure stripping model (see section~\ref{sec:rp_model}).  After matching the \citeauthor{yang2007} member galaxies to the SFR, stellar mass, and structural catalogues we are left with a total of 3250 galaxies in 24 clusters.
\par
We also make use of the isolated field sample from \citet{roberts2017} for comparison to the cluster sample.  The field sample is composed of all galaxies in single-member groups from the \citeauthor{yang2007} catalogue with a minimum separation of $1\,\mathrm{Mpc}$ and $1000\,\mathrm{km}\,\mathrm{s^{-1}}$ from their nearest `bright' neighbour.  Bright neighbours correspond to galaxies which are brighter than the $r$-band limiting absolute magnitude of the survey at $z = 0.1$, which ensures that the strictness of the isolation criteria is redshift independent.  Finally, any galaxies within $1\,\mathrm{Mpc}$ of the survey edge, or $1000\,\mathrm{km}\,\mathrm{s^{-1}}$ of the maximum redshift are removed.  Stellar masses and SFRs are obtained for the field sample from the same sources discussed above, resulting in an isolated, field sample of 164193 galaxies.

\section{SATELLITE QUENCHING VS. ICM DENSITY}
\label{sec:quench_dens}

\begin{figure*}
	\centering
	\includegraphics[width=0.7\textwidth]{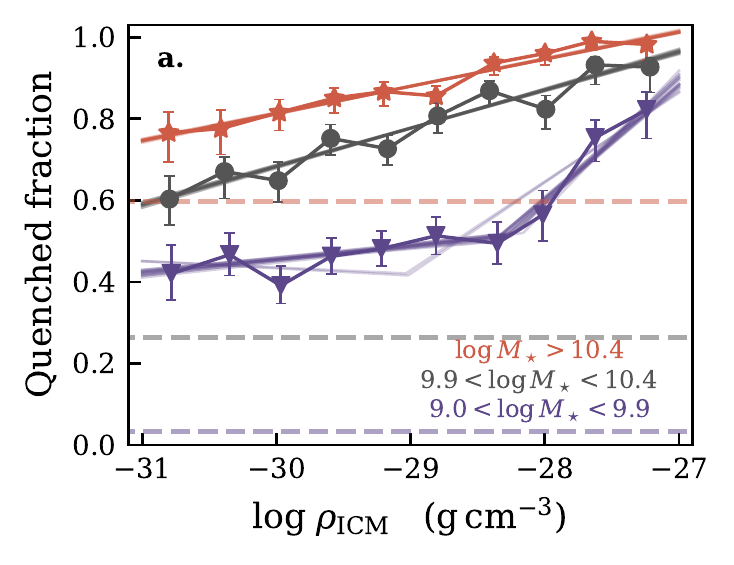}
	\caption{Quenched fraction ($\log\,\mathrm{sSFR} < -11\,\mathrm{yr^{-1}}$) versus ICM density for low- (purple), intermediate- (gray), and high-mass (red) galaxies.  Faded lines show fits to the data (single powerlaw for intermediate- and high-mass, double powerlaw for low-mass) after re-binning, ranging from 10 to 20 bins.  Quenched fractions for an isolated field sample in each mass bin are shown by the dashed horizontal lines.  Error bars correspond to 68 per cent Bayesian confidence intervals estimated from the beta distribution \citep{cameron2011}.  Uncertainty ranges on the field quenched fractions are smaller than the line widths shown in the figure.}
	\label{fig:qFrac}
\end{figure*}

It is well established by previous works that the fraction of quenched galaxies is a strong function of cluster-centric radius, from the local Universe out to at least $z \sim 1$ \citep{bamford2009, muzzin2012, presotto2012, wetzel2012, haines2015}.  Given the fact that there is significant scatter in the cluster-centric radius vs. ICM density relation across different clusters (see Fig.~\ref{fig:dens_profiles}), it is interesting to explore quenched fraction trends as a function of ICM density directly.  These trends can be used to explore quenching mechanisms such as ram pressure stripping (see section~\ref{sec:rp_comparison}), the strength of which depends explicitly on ICM density.  Fig.~\ref{fig:qFrac} shows quenched fraction ($\log\,\mathrm{sSFR} < -11\,\mathrm{yr^{-1}}$, \citealt{wetzel2013}) versus ICM density for low- (purple), intermediate- (gray), and high-mass (red) galaxies, computed in equally spaced bins of ICM density.  Galaxy stellar mass subsamples are defined such that there are an equal number of galaxies in each mass bin.  We also mark the quenched fraction for the isolated, field sample (described in section~\ref{sec:clusters}) in each mass bin as the dashed horizontal lines in Fig.~\ref{fig:qFrac}.
\par
We find that the quenched fraction at the lowest ICM densities, which corresponds to the vicinity of the virial radius ($R_{200m}$; equation~\ref{eq:Rvir}), is significantly enhanced relative to the value for the field, at the same stellar mass.  This is particularly true for the two lower mass bins, whereas the trend for the highest mass galaxies approaches the field value.  Numerous previous studies \citep[e.g.][]{lu2012, bahe2013, haines2015, roberts2017} have identified that the quenched fraction in the cluster outskirts can be significantly enhanced relative to the field, a fact which is often attributed to the pre-processing of star formation in less dense environments (ie.\ small groups) prior to infall onto galaxy clusters.  For the low-mass galaxies specifically, the data interpretted this way would require a pre-processed fraction of $\sim 30$ per cent.
\par
For intermediate- and high-mass galaxies the quenched fraction increases smoothly toward high ICM density. The environmental effect on the higher-mass galaxies is weaker than for the lowest-mass galaxies, consistent with previous studies that argue environment most strongly influences low-mass galaxies ($M_\star \lesssim \mathrm{few} \times 10^{10}\,\mathrm{M_\odot}$, \citealt{haines2006, bamford2009, peng2010}).  To ensure that the observed trends are not being driven by our particular binning scheme, we rebin the data from 10 bins up to 20 bins and fit the resulting trend for intermediate- and high-mass galaxies with a single powerlaw.  The fits to the 10 rebinnings are shown as the faded lines in Fig.~\ref{fig:qFrac}.  While increasing the number of bins adds noise and increases the statistical uncertainties, the fits remain nearly identical and the underlying trend is robust.  Given that the trend with ICM density is seemingly different for low-mass galaxies, these low-mass objects will be the focus of the remainder of the paper; however we will present a discussion of differences between high- and low-mass galaxies in Section~\ref{sec:discussion}.
\par
The quenched fraction trend for low-mass galaxies shows signs of a broken powerlaw, with a moderate increase at the lowest ICM densities and a steepening in the densest cluster regions.  To quantify this trend, we fit the quenched fraction trend for low-mass galaxies with both a single (SPL) and a broken powerlaw\footnote{http://docs.astropy.org/en/stable/modeling/} (BPL) and compare the best-fits.  Using the AIC as our model comparison tool, we find a lower value for the AIC for the broken powerlaw fit compared to the single powerlaw fit; this is true for each of the rebinnings (faded lines Fig.~\ref{fig:qFrac}).  This suggests that the AIC prefers a broken powerlaw, despite the penalty for increasing the number of fit parameters.  We note that this is not the case for the higher-mass bins, where a single powerlaw is always preferred. For the broken powerlaw the powerlaw slope is $\alpha_1 = 0.04_{-0.03}^{+0.01}$ at low ICM densities, and $\alpha_2 = 0.30_{-0.10}^{+0.09}$ at high ICM densities.  For the single powerlaw fit, the reduced chi-squared values range between 1.6 and 2.9 for the various rebinnings with a median of 2.2.  In comparison, the reduced chi-squared for the broken powerlaw ranges between 0.5 and 1.5 with a median of 0.8.  
\par
This apparent BPL trend for low-mass galaxies is an intriguing result, and if robust has important implications for the quenching of cluster satellites.  Further work is required to test the validity of this result.  We devote more time to the discussion of single versus double powerlaw along with other tests of robustness in section~\ref{sec:discussion}.
However, given the preference for the broken powerlaw according to the AIC and the reduced chi-squared, in sections~\ref{sec:rp_model} and \ref{sec:slow-then-rapid} we will take the broken powerlaw fit at face value in order to explore a potential origin for the shape of this trend and the implications for satellite quenching.

\section{ram pressure stripping model}
\label{sec:rp_model}

The broken powerlaw behaviour (for low-mass galaxies) in Fig.~\ref{fig:qFrac} matches the qualitative expectation for quenching via ram pressure stripping -- where a galaxy remains star forming until reaching a threshold density beyond which the ram pressure force becomes strong and quenching proceeds efficiently (see section~\ref{sec:rp_comparison}, Fig.~\ref{fig:stripFrac}, Fig.~\ref{fig:schematic}).  The break point in the broken power-law fit provides an estimate for this threshold density, and for our sample of low-mass galaxies we find a break point of $\log \rho_\mathrm{thresh} = -28.3_{-0.7}^{+0.2}\,\mathrm{g}\,\mathrm{cm^{-3}}$.  We now can test whether a simple, analytic ram pressure stripping model is able to reproduce the trend for low-mass galaxies in Fig.~\ref{fig:qFrac}.
\par
To directly constrain the fraction of galaxies susceptible to ram pressure stripping we take a simple analytic approach, similar to models used previously in literature \cite[e.g.][]{rasmussen2008, jaffe2015, jaffe2018}.  The basis of the model is the balance between ram pressure and the gravitational restoring force felt by the gas disc in a galaxy \citep{gunn1972}.  Specifically, gas will be susceptible to stripping when:
\begin{equation}\label{eq:rp}
\begin{split}
  \rho_\mathrm{ICM}(R)\,v^2 > [g_\mathrm{DM}(r) + & g_{d,\star}(r) + g_b(r) + \\
  g_\mathrm{HI}(r) + g_\mathrm{H_2}(r)]\,\Sigma_\mathrm{gas}(r)
\end{split}
\end{equation}
\noindent
where $\rho_{ICM}(R)$ is the density of the ICM as a function of cluster-centric radius, $v$ is the galaxy speed relative to the cluster centre, $\Sigma_\mathrm{gas}$ is the surface mass density of the atomic+molecular gas component, and $g(r)$ is the maximum restoring gravitational acceleration for the dark matter halo ($\mathrm{DM}$), the stellar disc ($d,\star$), bulge ($b$), atomic gas ($\mathrm{HI}$), and molecular gas ($\mathrm{H_2}$) in the direction perpendicular to the disc.  Given a model for the ICM density and the galaxy restoring force, the galactocentric radius at which stripping is efficient can be constrained.
\par
As described in section~\ref{sec:dens_profs}, we estimate the local ICM density for each galaxy using the beta profile fits to the deprojected \textit{Chandra} density profiles. To complete the left-hand-side of equation~\ref{eq:rp}, we estimate the galaxy speed relative to the cluster centre as,
\begin{equation}
  v = \sqrt{3} \times \frac{|z - z_\mathrm{cluster}|}{1 + z_\mathrm{cluster}} \times c
\end{equation}
\noindent
where $z$ is the galaxy redshift, $z_\mathrm{cluster}$ is the cluster redshift from the Yang catalogue, and $c$ is the speed of light.  The factor of $\sqrt{3}$ is included to convert from line-of-sight to three dimensional speed, on average.  Following equation~\ref{eq:rp}, the restoring gravitational accelerations for both the gas and stellar components have to be modeled for each satellite galaxy as a function of galactocentric radius.  For the stellar distribution the bulge and disc components are modelled with GIM2D bulge + disc decompositions (exponential disc, De Vaucouleurs bulge; \citealt{simard2011}), giving disc scale lengths and bulge effective radii ($R_d,\;R_e$) in the $r$-band.  We also make use of bulge and disc stellar masses ($M_b,\;M_d$) from \citet{mendel2014}.  For the stellar disc, we assume an exponential profile and calculate the restoring gravitational acceleration as
\begin{equation}\label{eq:g_disc}
  g_{d,\star} = 2 G \Sigma_{d,\star}(r) = 2 G \Sigma_0 \, e^{-r / R_d}
\end{equation}
\noindent
where $\Sigma_{d,\star}(r)$ is the surface density of the stellar disk as a function of radius.  Given the disc mass and scale length, the normalization, $\Sigma_0$, can be determined by integrating the surface density,
\begin{equation}
  M_d = 2 \pi \int_0^\infty \Sigma_{d,\star}(r) r\,\mathrm{d}r = 2 \pi \Sigma_0 R_d^2.
\end{equation}
\noindent
For the bulge component, we assume a Hernquist profile, as it has a convenient analytic form which is a good approximation to the De Vaucouleurs profile assumed in the bulge+disc decomposition \citep{hernquist1990}.  The bulge potential, $\phi_b$, is then given by
\begin{equation}
  \phi_b(r,z) = -\frac{G M_b}{(r^2 + \mathcal{Z}^2)^{1/2} + a}
\end{equation}
\noindent
where $\mathcal{Z}$ corresponds to the distance in direction perpendicular to the disc component and $a$ is related to the bulge effective radius as $a = R_e / 1.815$ \citep{hernquist1990}.  We then calculate the maximum (over the $\mathcal{Z}$ direction) restoring gravitational acceleration as
\begin{equation}
\begin{split}
  g_b(r) &= \max_\mathcal{Z} \frac{\partial \phi_b(r,\mathcal{Z})}{\partial \mathcal{Z}} \\
  &= \max_\mathcal{Z} \frac{G M_b}{[(r^2 + \mathcal{Z}^2)^{1/2} + a]^2}\frac{\mathcal{Z}}{(r^2 + \mathcal{Z}^2)^{1/2}}.
\end{split}
\end{equation}
\noindent
For the dark matter halo, we also assume a Hernquist profile.  The galaxy halo mass is determined using the observed stellar mass and the stellar-to-halo mass relation from \citet{hudson2015}.  The scale radius, $a = r_\mathrm{vir} / c$, is estimated from the concentration and the galaxy $r_{200}$, assuming a concentration-mass relation from \citet{diemer2015}.  Both the bulge and dark matter halo profiles are implemented using the \textsc{colossus}\footnote[2]{https://bdiemer.bitbucket.io/colossus/} package \citep{diemer2017}.

\begin{figure}
	\centering
	\includegraphics[width=\columnwidth]{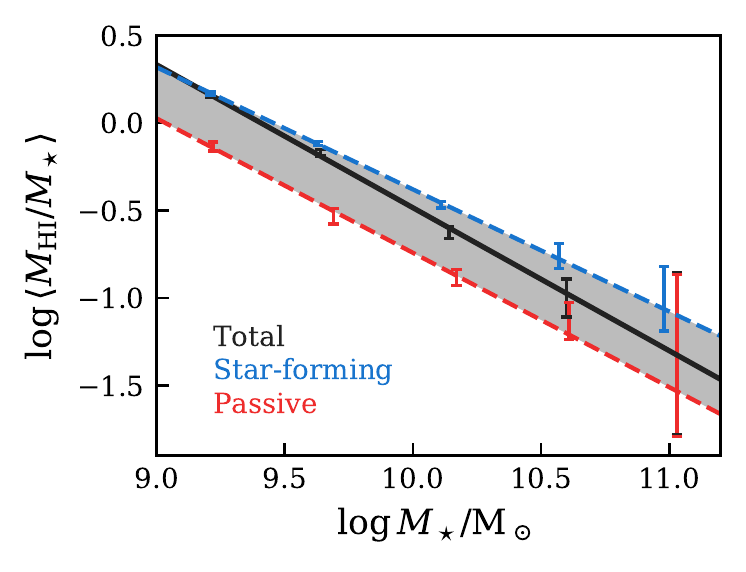}
	\caption{HI gas fraction -- stellar mass scaling relation obtained by stacking ALFALFA spectra (both detections and non-detections) for 14128 low redshift SDSS galaxies.  Scaling relations are shown for the total sample (black), as well as just star-forming (blue) and passive (red) galaxies.}
	\label{fig:fgas_m}
\end{figure}

Modeling the gas component is more challenging as, unlike the stellar component, these galaxies are generally not observed in atomic or molecular gas.  We assume that the gas distribution consists of an atomic and molecular component, each following an exponential disc with different scale lengths -- with the restoring acceleration taking the same exponential form as equation~\ref{eq:g_disc}.  We assume that the atomic gas scale length is twice the optical disc scale length (ie.\ $R_\mathrm{HI} = 2\,R_d$), which is consistent with the observed value for local non-HI deficient galaxy discs \citep{cayatte1994, cortese2010, boselli2014}.  For the molecular component we assume a scale length equal to the optical disc scale length (ie.\ $R_\mathrm{H_2} = R_d$), again corresponding to molecular gas-to-stellar size ratio for non-deficient galaxy discs \citep{boselli2014}.  Because the galaxies in this sample are not observed in atomic or molecular gas, we assign gas masses statistically using the atomic/molecular gas fraction - stellar mass ($f_\mathrm{gas}$ vs. $M_\star$, $f_\mathrm{gas} = M_\mathrm{gas} / M_\star$) scaling relations for representative samples of galaxies.  In particular, we assign HI gas fractions for a given stellar mass assuming the $f_\mathrm{HI} - M_\star$ relation for isolated galaxies.  We are therefore approximating the gas fraction that the cluster galaxies in this sample would have had in the field, prior to infall; allowing us to estimate the amount of pre-infall gas that can be stripped by the cluster environment.  To do this we construct the $f_\mathrm{HI} - M_\star$ relation using the spectral stacking technique \citep{fabello2011} applied to a sample of 14128 isolated, low-redshift ($0.02 < z < 0.05$) SDSS galaxies with processed ALFALFA data \citep{brown2017}.  We direct the reader to \citet{brown2015} for a complete description of the parent sample and stacking technique, however most importantly, spectral stacking allows us to exploit both HI detections and non-detections to obtain the $f_\mathrm{HI} - M_\star$ relation for an unbiased sample of galaxies from low ($\sim 10^9\,\mathrm{M_\odot}$) to high ($\sim 10^{11}\,\mathrm{M_\odot}$) stellar mass.  Fig.~\ref{fig:fgas_m} shows the $f_\mathrm{HI} - M_\star$ relation for the sample of isolated SDSS galaxies for the total sample (black) as well as for subsamples of star-forming (blue, 10984 galaxies with $\log\,\mathrm{sSFR} > -11\,\mathrm{yr^{-1}}$) and passive (red, 3144 galaxies $\log\,\mathrm{sSFR} < -11\,\mathrm{yr^{-1}}$) galaxies.  A consequence of the spectral stacking technique is that the intrinsic scatter in the $f_\mathrm{HI} - M_\star$ relation is unknown, therefore we use the difference between the relation for star-forming and passive galaxies as a rough estimate of the scatter.  We assign `pre-infall' HI masses to each cluster galaxy by stochastically sampling from the `scatter' in Fig.~\ref{fig:fgas_m} (the shaded region) at the stellar mass of the galaxy.  The stochastic selection is weighted by the inverse distance from the relation for the total sample (black line), therefore the sampling reflects the fact the scatter is not symmetric about the trend for the total sample.  This process is iterated 1000 times and final values calculated using the HI mass are taken to be the median of these Monte Carlo re-samplings.
\par
Pre-infall molecular gas masses ($M_{\mathrm{H_2}}$) are assigned using the $f_{\mathrm{H_2}} - M_\star$ relation from  xCOLD GASS \citep{saintonge2017} obtained from spectral stacking using a representative sample of low redshift galaxies ($0.01 < z < 0.02$, $M_\star > 10^9\,\mathrm{M_\odot}$).  \citet{saintonge2017} find that, for the stacked sample, the molecular gas fraction is approximately constant at $f_{\mathrm{H_2}} \sim 0.1$ for masses $\lesssim 10^{10.5}\,\mathrm{M_\odot}$ (see table 5 in \citealt{saintonge2017}).  For the ram-pressure model we focus on low-mass galaxies ($M_\star < 10^{10}\,\mathrm{M_\odot}$), therefore we choose to assign a constant molecular gas fraction of $f_{\mathrm{H_2}} = 0.1$ to our cluster galaxies.  We note that at these masses the gas content of galaxies remains largely dominated by the atomic component \citep[e.g.][]{saintonge2017}, which is also the component most suceptible to environmental interactions.  Therefore our specific assumptions regarding the molecular gas component do not strongly affect the results.
\par
Using equation~\ref{eq:rp}, combined with the galaxy and ICM models described above we can determine the stripping radius, $r_\mathrm{strip}$, the galactocentric radius at which $\rho_\mathrm{ICM}(R)\,v^2 > [g_\mathrm{DM}(r) + g_{d,\star}(r) + g_b(r) + g_\mathrm{HI}(r) + g_\mathrm{H_2}(r)] \Sigma_\mathrm{gas}(r)$. Given the stripping radius, the stripped mass (ie.\ the gas mass outside of the stripping radius) can be calculated as,
\begin{equation}
  M_\mathrm{strip} = M_\mathrm{gas}\,e^{-r_\mathrm{strip} / R_\mathrm{gas}}\,\left(\frac{r_\mathrm{strip}}{R_\mathrm{gas}} + 1 \right)
\end{equation}
\noindent
 \citep{BT2008}.  We compute the stripped mass for both the atomic and molecular gas components and then calculate the total stripped gas fraction as,
\begin{equation}
  f_\mathrm{strip} = \frac{M_\mathrm{strip,\,HI} + M_\mathrm{strip,\,\mathrm{H_2}}}{M_\mathrm{HI} + M_\mathrm{\mathrm{H_2}}}.
\end{equation}
\noindent
Uncertainties on the disk scale lengths and bulge effective radii from \citet{simard2011} introduce uncertainties on the stripped masses of roughly 10 per cent, but do not bias the results in any way.
\par
We stress that the ram pressure stripping model described above makes many simplifying assumptions.  This model is not intended to be a detailed treatment of ram pressure stripping, but instead to provide a rough estimate of where ram pressure is expected to significantly influence satellite galaxies.  We highlight and discuss the primary assumptions of this model in appendix~\ref{sec:model_appendix}.

\subsection{Comparison to observed quenched fractions}
\label{sec:rp_comparison}

\begin{figure*}
	\centering
	\includegraphics[width=0.7\textwidth]{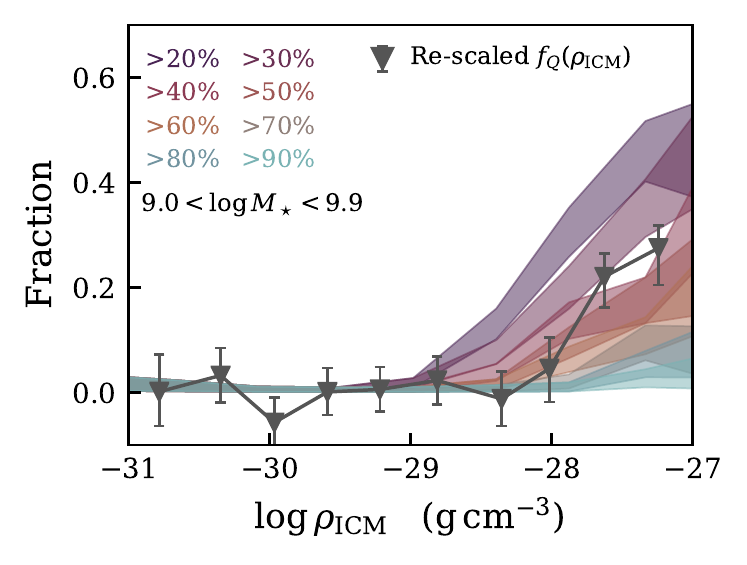}
	\caption{Fraction of low-mass galaxies with the listed percentage of their gas mass susceptible to stripping, as a function of ICM density.  Shaded bands correspond to 68 per cent Bayesian confidence intervals estimated from the beta distribution \citep{cameron2011}.  For comparison to the shape of these tracks we plot the quenched fraction for low-mass galaxies from Fig.~\ref{fig:qFrac} with the powerlaw fit at low ICM density subtracted off, this permits a direct comparison between ram pressure stripping and the observed quenched fraction upturn at high ICM density.}
	\label{fig:stripFrac}
\end{figure*}

Using this analytic model of ram pressure stripping, we can now make direct comparisons to the observed quenched fractions for the low-mass galaxies in Fig.~\ref{fig:qFrac}.  Specifically, we aim to constrain whether the apparent quenched fraction upturn at high ICM density and low stellar mass can be reproduced by a simple ram pressure model.  Therefore, we isolate the high density upturn by subtracting off the powerlaw fit to the low ICM density trend ($\rho_\mathrm{ICM} \lesssim 10^{-28}\,\mathrm{g}\,\mathrm{cm^{-3}}$).  This allows us to make a direct comparison between the observed upturn and the output from our ram pressure model.  We calculate the fraction of galaxies in the sample which have $n$-per cent of their cold gas reserves susceptible to stripping (where $n$ is a free parameter in the model).
\par
In Fig.~\ref{fig:stripFrac} we overlay the re-scaled quenched fraction, with the low density powerlaw subtracted off, on top of output tracks from the ram pressure model.  Each track corresponds to the fraction of galaxies in the sample which have \textit{at least} the given percentage of their cold-gas mass located beyond the stripping radius (and therefore susceptible to stripping).  We show tracks ranging between $>$20 and $>$90 per cent and the observed quenched fraction trend is well matched by a model where $\sim$half of a galaxy's cold-gas mass is available to be stripped.  The stripped fractions in Fig.~\ref{fig:qFrac} correspond to the removal of a parcel of gas entirely from the galaxy, as they are computed using the maximum restoring accelerations (in the direction perpendicular to the disc, equation~\ref{eq:rp}).  Considering instead the removal of gas from a typical gaseous disc (scale height $\sim$ few hundred parsecs), the data are then well fit by a stripped fraction of $\gtrsim 70$ per cent.  Regardless of the precise definition of gas stripping, we emphasize that the primary takeaway is the fact that the observed upturn in quenched fraction coincides closely to the onset of a significant ram pressure force relative to the galaxy restoring potential.  The results from this simplified ram pressure model are consistent with low-mass cluster galaxies experiencing enhanced quenching due to relatively efficient ram pressure.  Below this threshold density, ram pressure seems to be not strong enough, on average, to quench even low-mass galaxies.

\section{Slow-then-rapid quenching}
\label{sec:slow-then-rapid}

\begin{figure*}
	\centering
	\includegraphics[width=\textwidth]{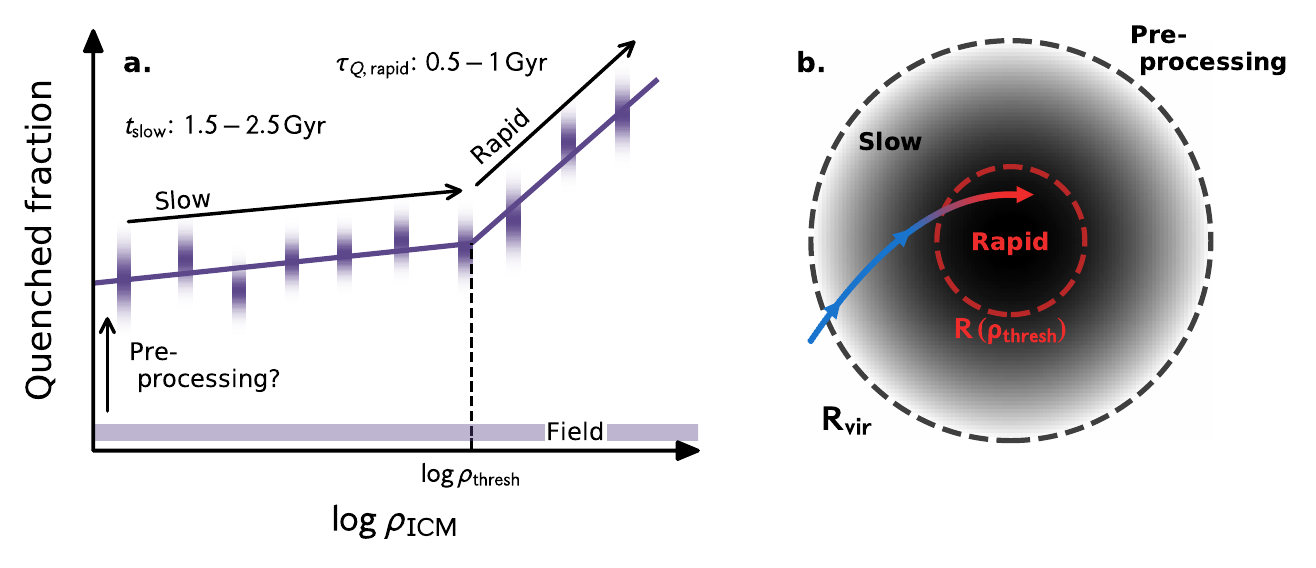}
	\caption{Schematic diagram illustrating the connection between the results of this paper and a ``slow-then-rapid'' quenching model.  At galaxy infall (at the virial radius), the quenched fraction for infalling galaxies is significantly larger than the corresponding value for isolated, field galaxies, at the same stellar mass.  This offset between the infalling and field populations is consistent with the pre-processing of star formation prior to infall.  The slow-quenching phase occurs after galaxy infall prior to reaching a threshold ICM density. Galaxies spend $1.5-2.5\,\mathrm{Gyr}$ between infall and reaching the threshold ICM density.  At a threshold ICM density ($\rho_\mathrm{ICM} \sim 10^{-28}\,\mathrm{g}\,\mathrm{cm^{-3}}$), a significant fraction of a galaxy's cold gas mass is now susceptible to ram pressure stripping and quenching can occur rapidly on an e-folding time $\tau_Q \sim 1\,\mathrm{Gyr}$, at ICM densities higher than the threshold value.  The left-hand panel annotates this interpretation on the observed quenched fraction versus ICM density plot, where each data point is shown a Gaussian `smear' indicating the uncertainty.  The right-hand panel illustrates this model showing a hypothetical infall track of a star-forming galaxy onto a cluster.}
	\label{fig:schematic}
\end{figure*}

A commonly invoked model for environmental quenching is so-called delayed-then-rapid quenching \citep{wetzel2013}, where satellite quenching does not occur immediately upon infall but proceeds rapidly only after a characteristic delay time.  Qualitatively, the low-mass quenched fraction versus ICM density trend found in this work lends itself naturally to a similar interpretation (see Fig.~\ref{fig:qFrac}), namely a ``slow-then-rapid'' quenching framework.  Again taking the apparent BPL trend at face value (see section~\ref{sec:discussion} for a detailed discussion of the robustness of this trend), we interpret the modest powerlaw slope between infall and a galaxy reaching the threshold ICM density as the slow-quenching portion, beyond which point the galaxy rapidly quenches as it moves to higher ICM density.  This interpretation is illustrated schematically in Fig.~\ref{fig:schematic} both as an annotated quenched fraction plot, and as a diagram showing a toy infall track for a star-forming galaxy onto a cluster.
\par
To be more quantitative, and to permit comparisons to previous studies, we derive rough estimates for the time that a satellite galaxy spends on the slow-quenching track after infall ($t_\mathrm{slow}$) as well as the quenching timescale associated with the rapid-quenching component ($\tau_{Q,\mathrm{rapid}}$) using a simple exponential model.  The primary simplifying assumption made in this model is that galaxies are quenched exclusively on their first infall on a radial orbit.  We note that these assumptions are consistent with recent results from hydrodynamic simulations \citep{lotz2018, arthur2019}.
\par
We calculate the time spent in the slow-quenching mode as,
\begin{equation}\label{eq:slow_t}
  t_\mathrm{slow} = \left(\frac{R_{200m} - R_\mathrm{thresh}}{v_\mathrm{slow}}\right)
\end{equation}
\noindent
where $R_{200m} - R_\mathrm{thresh}$ is the radial distance traveled between infall and reaching the threshold ICM density, and $v_\mathrm{slow}$ is the mean galaxy velocity over the `slow-quenching' portion.  For each cluster, $R_{200m}$ is given by equation~\ref{eq:Rvir}, and we measure $R_\mathrm{thresh}$ given the observed density profile (Fig.~\ref{fig:dens_profiles}).  To determine galaxy velocities we make use of the high-resolution dark matter only simulations from \cite{joshi2016}.  Specifically, we consider galaxy cluster subhalos on their first infall, located between $R_{200m}$ and the median $R_\mathrm{thresh}$ for the clusters in our sample, $0.25 \times R_{200m}$.  We only consider subhalos in $>10^{14}\,\mathrm{M_\odot}$ clusters with peak dark matter masses between $10^{11.1} < M_\mathrm{peak} < 10^{11.75}\,\mathrm{M_\odot}$ which corresponds to our low-mass galaxy stellar mass range of $10^9 \lesssim M_\star \lesssim 10^{10}\,\mathrm{M_\odot}$ assuming a stellar-to-halo mass relation from \citet{hudson2015}.  We normalize the subhalo velocities by the one-dimensional velocity dispersion of the host clusters in the $z = 0$ snapshot, which is directly comparable to the measured line-of-sight velocity dispersions for the clusters in our sample.  The cluster velocity dispersions measured from the dark matter simulations and the velocity dispersions measured for the observed clusters are both calculated using the biweight estimator \citep{beers1990}.  For subhalos on first infall, located between $0.25 < R_\mathrm{3D} < 1\,R_{200m}$, we find the median velocity is $v = 1.8\times \sigma_\mathrm{1D}$.  To account for the spread in infall velocities, we calculate $t_\mathrm{slow}$ in a Monte Carlo sense by sampling velocities, $v_\mathrm{slow}$, from the full distribution extracted from the simulations.  We measure $t_\mathrm{slow}$ as the median of 1000 random samplings.  Using equation~\ref{eq:slow_t}, we obtain an estimate of $t_\mathrm{slow} = 1.8_{-0.3}^{+0.6}\,\mathrm{Gyr}$.  Therefore, infalling galaxies spend $\sim1.5-2.5\,\mathrm{Gyr}$ before reaching the threshold ICM density, beyond which quenching proceeds rapidly.  The value for $t_\mathrm{slow}$ is, unsurprisingly, close to the dynamical time for cluster-mass halos.
\par
For the rapid portion we model the quenched fraction as increasing exponentially over a characteristic e-folding time, $\tau_Q$.  The quenched fraction is then given by,
\begin{equation}
  f_Q = 1 - (1 - f_{Q,0})\,e^{-t / \tau_Q}.
\end{equation}
\noindent
Specifically, the e-folding timescale for rapid-quenching is estimated as
\begin{equation}\label{eq:rapid_tau}
  \tau_{Q,\mathrm{rapid}} = t_\mathrm{rapid} \times \left[\ln \left(\frac{1}{1 - f_{Q,\mathrm{rescaled}}(\rho^\prime)}\right)\right]^{-1}
\end{equation}
\noindent
with,
\begin{equation}\label{eq:rapid_t}
  t_\mathrm{rapid} = \left(\frac{R_\mathrm{thresh} - R(\rho^\prime)}{\bar{v}_\mathrm{rapid}}\right).
\end{equation}
\noindent
We take $\rho^\prime = 10^{-27}\,\mathrm{g}\,\mathrm{cm^{-3}}$ and $f_{Q,\mathrm{rescaled}}(\rho^\prime) = 0.35$, where $f_{Q,\mathrm{rescaled}}$ is the rescaled quenched fraction where the contribution from the low-ICM density powerlaw has been subtracted in order to isolate the high-density upturn (see Fig.~\ref{fig:stripFrac}).  The characteristic infall velocity is again determined from dark matter simulations, now for subhalos at $R_\mathrm{3D} < 0.25 \times R_{200m}$.  In this inner cluster region we find the median velocity $v = 2.8 \times \sigma_\mathrm{1D}$, and again estimate the timescale by randomly sampling the simulated velocity distributions.  Using equation~\ref{eq:rapid_tau}, this gives a median quenching e-folding time of $\tau_{Q,\mathrm{rapid}} = 0.6_{-0.1}^{+0.1}\,\mathrm{Gyr}$.
\par
Our estimates for $t_\mathrm{slow}$ and $\tau_{Q,\mathrm{rapid}}$ suggest a total quenching time of $\sim 2-3\,\mathrm{Gyr}$ for low-mass galaxies in clusters.  Studying cluster dwarf galaxies in the Illustris simulation, \citet{mistani2016} measure the time ellapsed between infall and the first time a galaxy's sSFR falls below $10^{-11}\,\mathrm{yr^{-1}}$, finding timescales ranging between $\sim 3 - 5.5\,\mathrm{Gyr}$ for stellar masses $10^9 \lesssim M_\star \lesssim 10^{10}\,\mathrm{M_\odot}$.  \citet{haines2015} employ a model where galaxies in LoCuSS clusters are quenched instantaneously after a delay time $\Delta t$ since infall, and find that the surface density of star-forming galaxies based on infrared (UV) observations is best-fit by a delay time of $2.1_{-0.7}^{+0.8}$ ($3.2 \pm 0.4$) Gyr.  \citet{haines2015} also derive quenching times of $\sim 1.5 - 2\,\mathrm{Gyr}$ assuming that star formation declines exponentially upon cluster infall.  \citet{wetzel2013} obtain total quenching times of $4-5\,\mathrm{Gyr}$ for low-mass satellites of cluster-mass halos, which are somewhat larger than the estimates derived in this work.  Using semi-analytic models applied to the Millenium Simulation \citep{springel2005} along with observations of galaxies in groups and clusters from the \citet{yang2007} catalogue, \citet{delucia2012} argue that on average galaxies spend $5 - 7\,\mathrm{Gyr}$ in halos $>10^{13}\,\mathrm{M_\odot}$ before quenching.  These timescales are longer than the total quenching timescale that we derive, however given that our sample is dominated by large clusters (median halo mass, $4 \times 10^{14}\,\mathrm{M_\odot}$) it is difficult to make a direct comparison.  In fact we do see evidence for pre-processing, meaning that many galaxies in our cluster sample were likely members of smaller groups ($\sim 10^{13}\,\mathrm{M_\odot}$) prior to cluster infall.  Including potential additional time spent as satellites of smaller groups could bring our quenching timescales closer to the \citet{delucia2012} estimates.  In general, the quenching times that we derive for low-mass cluster galaxies are roughly consistent, if somewhat smaller, than previous estimates from the literature.

\section{DISCUSSION}
\label{sec:discussion}

\subsection{Pre-processing}

In Fig.~\ref{fig:qFrac} we show that the quenched fraction in the cluster outskirts is significantly enhanced relative to the field, especially for low-mass galaxies.  A natural interpretation for this result is that a fraction of infalling galaxies have been pre-processed prior to infall onto the current cluster.  The data for low-mass galaxies require a pre-processed fraction of $\sim 0.3$ to account for the difference from the field value.  This pre-processed fraction estimate, however, is quite crude and does not account for any contamination from backsplashing galaxies which will have the effect of artificially increasing the apparent level of pre-processing.  Therefore, it is more precise to treat this value as an upper limit.  With that in mind, this fraction is roughly consistent with estimates for the pre-processed fraction from previous studies.  For example, estimates of the fraction of galaxies which infall onto clusters as members of smaller groups (where the galaxies would be susceptible to pre-processing) range from $\sim\! 25$ to $\sim\! 60$ per cent using both simulations \citep{mcgee2009, delucia2012, bahe2013} and observations \citep{hou2014}.  These constraints provide an upper limit to the pre-processed fraction, as not all galaxies infalling as a member of a group will necessarily have been quenched.  More direct constraints on the fraction of pre-processed galaxies range from $\sim\!10-30$ per cent for cluster galaxies \citep{haines2015, roberts2017, vanderburg2018}.  The fraction derived in this work falls on the upper of this range, however, it is again important to note that a portion of the satellite population near the virial radius are actually backsplash galaxies that have already made a pericentric passage \citep[e.g.][]{mahajan2011, bahe2013, oman2013, hirschmann2014}.  Therefore it is likely that quenched backsplash galaxies at the virial radius are contaminating the `infalling' population, and artificially increasing the apparent pre-processed fraction.
\subsection{Is a broken powerlaw required for the low-mass data?}
\label{sec:pl_discussion}

\begin{figure*}
	\centering
	\includegraphics[width=\textwidth]{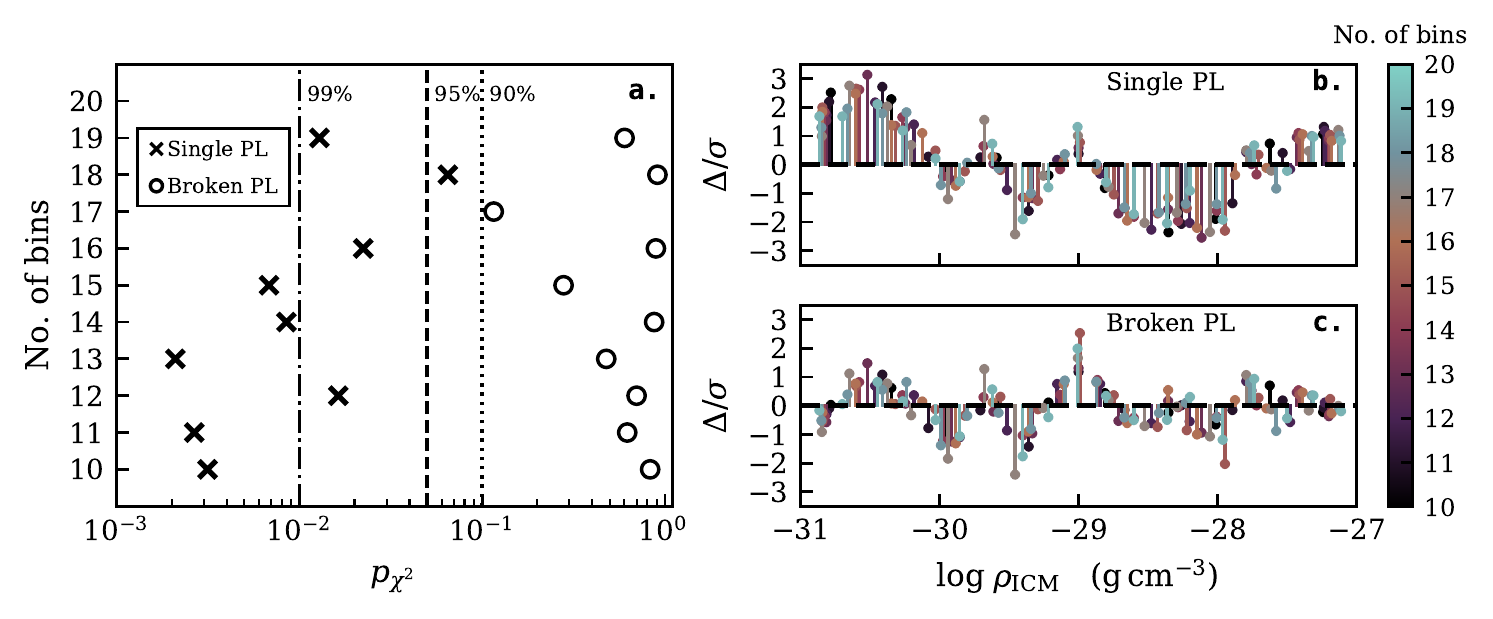}
	\caption{\textit{Left:} P-value from chi-squared test for various re-binnings (number of bins, y-axis), for single (crosses) and double (circles) powerlaw fits.  \textit{Top-right:} Residuals from the single powerlaw fit for each of the re-binnings (see colourbar for number of bins).  \textit{Bottom-right:} Residuals from the broken powerlaw fit for each of the re-binnings (see colourbar for number of bins).  Residuals are normalized by the 68 per cent uncertainty on data points.}
	\label{fig:fit_cs}
\end{figure*}

A key question to address in this study is not only whether the BPL gives a better fit than the SPL to the low-mass data, but also whether or not the SPL provides an acceptable fit (regardless of the quality of the BPL fit).  We have already made use of the AIC as a model discriminator, and as is discussed in section~\ref{sec:quench_dens}, the AIC favours the BPL fit over the SPL fit for the low-mass data and all of the re-binnings of these data.  This is certainly clear evidence that the BPL fit is statistically preferred over the simpler SPL, however the AIC says nothing about the quality of individual fits, only whether one fit is better than another.
\par
We also test the significance of the single and broken powerlaw fits individually by applying a simple chi-squared test.  In Fig.~\ref{fig:fit_cs} we plot the p-value from the chi-squared test for both the single (crosses) and broken (circles) powerlaw fits for each of the re-binnings, ranging from 10 to 20 bins on the y-axis.  For all of the re-binnings we see evidence (at the 90 per cent level) that the SPL does not provide a sufficient fit to describe the low-mass galaxy trend.  This suggests that the SPL fit may not be sufficient based on the chi-squared test.  One aspect of the fit that a simple chi-squared test does not take into account is any structure in the fit residuals.  In Figs~\ref{fig:fit_cs}b and \ref{fig:fit_cs}c we show the fit residuals (normalized by the uncertainty on each data point) for the single and double powerlaw fits, respectively.  We show the residuals for the fits to all of the re-binnings and mark the number of bins in each fit with the colourbar in Figs~\ref{fig:fit_cs}b and \ref{fig:fit_cs}c.  Two trends are clear by comparing Figs~\ref{fig:fit_cs}b and \ref{fig:fit_cs}c:  1. The amplitude of the residuals are smaller for the BPL fit relative to the SPL.  2.  For the BPL fit the residuals seem to be randomly scattered around the zero-line, whereas there is apparent structure in residuals for the SPL fit with a consistent excess of positive residuals between $10^{-31}$ and $10^{-30}\,\mathrm{g}\,\mathrm{cm^{-3}}$ and an excess of negative residuals between $10^{-29}$ and $10^{-28}\,\mathrm{g}\,\mathrm{cm^{-3}}$.  This structure in the residuals for the SPL fit is evidence that the SPL model does not fully describe the low-mass quenched fraction data.

\subsection{Robustness tests for the broken powerlaw trend}

\subsubsection{SFR indicator}

\begin{figure}
	\centering
	\includegraphics[width=\columnwidth]{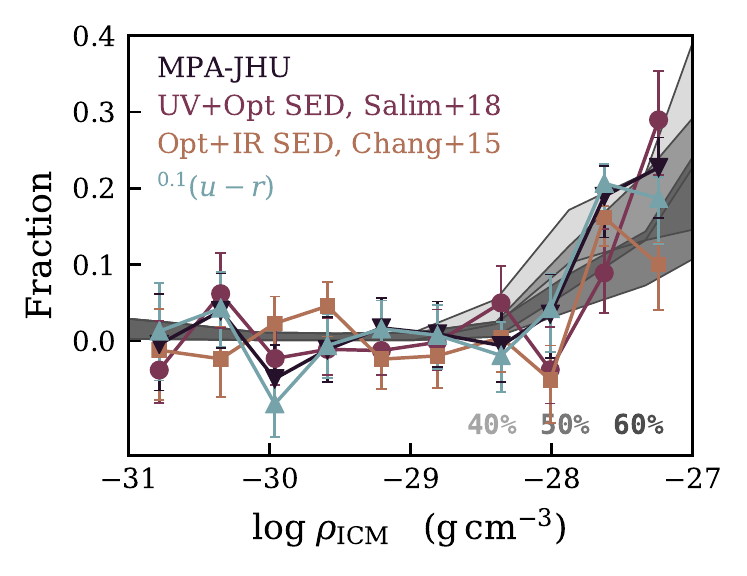}
	\caption{Baseline subtracted quenched fraction as a function of ICM density for four different star formation rate tracers: $\mathrm{H\alpha}+\mathrm{D_n4000}$ (downward triangles, \citealt{brinchmann2004}), $u-r$ colour (upward triangles, \citealt{blanton2007}), UV+Optical SED fitting (circles, \citealt{salim2016, salim2018}), Optical+IR SED fitting (squares, \citealt{chang2015}).  In the background we plot our analytic ram pressure tracks for $>$40, $>$50, and $>$60 per cent stripped.  The characteristic break at $\rho_\mathrm{ICM} \simeq 10^{-28}\,\mathrm{g}\,\mathrm{cm^{-3}}$ is present regardless of star formation rate tracer.}
	\label{fig:sfr_tracers}
\end{figure}

When calculating quenched fractions we use the MPA-JHU SFRs, which are derived from $\mathrm{H\alpha}$ emission (when detected; \citealt{brinchmann2004}).  $\mathrm{H\alpha}$ has the advantage of being a tracer of star formation on very short timescales ($\lesssim 10\,\mathrm{Myr}$, \citealt{kennicutt2012}) which is especially important when investigating rapid quenching mechanisms (such as ram pressure).  At low SFRs (where emission lines are not detected) the MPA-JHU values are based on the $\mathrm{D_n4000}$ break and are therefore less precise.  Given the large number of value-added catalogues available for the Sloan Digital Sky Survey, we are able to reproduce Fig.~\ref{fig:stripFrac} for different star formation rate estimators to test whether the observed shape is driven by our choice of star formation indicator.  In Fig.~\ref{fig:sfr_tracers} we show the re-scaled quenched fraction (where we have subtracted the low density powerlaw, as in Fig.~\ref{fig:stripFrac}) versus ICM density for SFRs derived from UV+Optical+mid IR SED fitting \citep{salim2016, salim2018}, Optical+IR SED fitting \citep{chang2015}, rest-frame\footnote[2]{k-corrected to $z=0.1$} $u-r$ colours \citep{blanton2007} (where we assume that the red fraction corresponds to the quenched fraction), along with the primarily $\mathrm{H\alpha}$ SFRs \citep{brinchmann2004} from the main text for reference.  For all SFR estimators we define quenched galaxies to have $\mathrm{sSFR} < 10^{-11}\,\mathrm{yr^{-1}}$, and for $u-r$ colours we define quenched galaxies to have $^{0.1}(u-r) > 2.4$ which corresponds to the intersection between the red sequence and blue cloud using a double-gaussian fit.  Also plotted for reference are the ram pressure stripping tracks for stripped fractions of $>$40, $>$50, and $>$60 per cent.  Fig.~\ref{fig:sfr_tracers} shows that the general trend presented in sections~\ref{sec:quench_dens} \& \ref{sec:rp_comparison} persists regardless of SFR indicator, and that in all cases the trend is well matched by a model where quenching becomes efficient when $\sim$half of the galactic cold-gas reservoir is susceptible to ram pressure stripping.

\subsubsection{ICM density uncertainty at large radius}
\label{sec:dens_steep}

\begin{figure}
	\centering
	\includegraphics[width=\columnwidth]{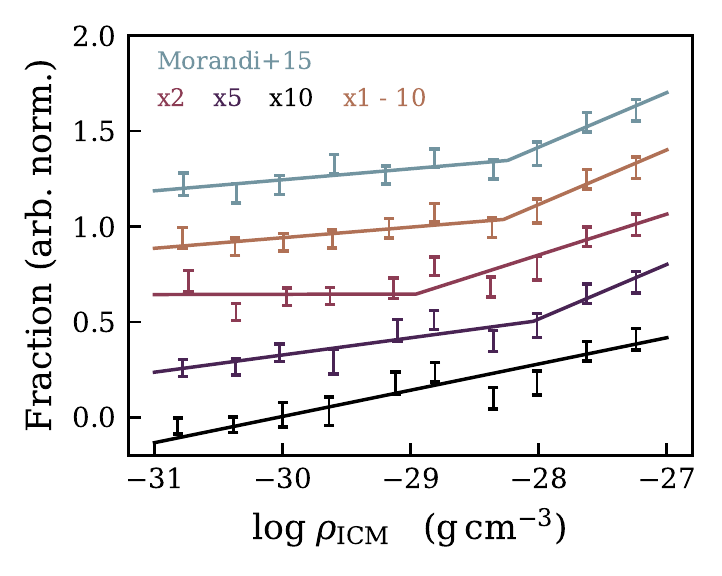}
	\caption{Quenched fraction (each offset by 0.3 in the vertical direction) vs. ICM density for various assumptions about ICM density beyond $R_{500}$.  `x2', `x5', and `x10' correspond to systematically decreasing the ICM density estimate by the given factor.  `x1 - 10' corresponds to randomly decreasing the ICM density for each galaxy by a factor between one and ten.  `Morandi+15' corresponds to steepening the observed gas density profile slope by a factor between 1 and 1.75 (taken from the observations in \citealt{morandi2015}) according to $R/R_{500}$ for each galaxy.}
	\label{fig:dens_shift}
\end{figure}

Given that the X-ray data used in this work only reach sufficient depth ($>5000$ counts per annulus) within $R_{500}$, it is necessary to extrapolate the density profiles to obtain local ICM density estimates for galaxies in the cluster outskirts.  Previous work has shown that ICM density profiles tend to steepen at large radius ($\gtrsim R_{500}$) compared to the inner regions \citep{morandi2015}.  In addition, clumpy gas distributions are common in the cluster outskirts and can introduce biases affecting estimates of the gas density \citep[e.g.][]{walker2013, morandi2014, ichinohe2015}.  This uncertainty in the ICM density profiles at large radius is an important source of uncertainty for this analysis, however it is difficult to quantify on a case-by-case basis without deep X-ray observations out to the virial radius.
\par
Given the observed steepening of density profiles at large radii, by extrapolating profiles we may be overestimating the ICM density in the cluster outskirts.  We employ a few different methods to test what effect this could have on the broken power-law trend we observe.  As a simple first test we arbitrarily decrease the local ICM density estimate for each galaxy beyond $R_{500}$ by a constant factor.  The lines labelled `x2', `x5', and `x10' in Fig.~\ref{fig:dens_shift} correspond to the quenched fraction trend after assuming a decrease in ICM density by a factor of two, five, and ten beyond $R_{500}$.  Note that lines in Fig.~\ref{fig:dens_shift} have been offset by 0.3 in the vertical direction for readability. For a decrease of a factor of two and five the broken power-law shape is still evident (and is still preferred over a single power-law by the AIC), however for a factor of ten decrease there is no evidence for a broken power-law trend.  Therefore unless we are systematically overestimating the local ICM density for each galaxy beyond $R_{500}$ by a factor of $\ge10$, the broken power-law trend appears robust.  The line labelled `x1-10' in Fig.~\ref{fig:dens_shift} corresponds to decreasing the local ICM density randomly by a factor between one and ten.  Specifically, the plotted line shows the median of 1000 random Monte Carlo trials, and the presence of the broken power-law shape is evident and confirmed by the AIC.  Finally, \citet{morandi2015} study a large sample ($>$300) of galaxy clusters observed by \textit{Chandra} with coverage out to and beyond the virial radius.  For their low-redshift sample ($z < 0.3$) they find that the slope of the gas density profile ($\beta$) steadily steepens by a factor of 1.75 between $R_{500}$ and $R_{100}$ ($R_{100}$ is similar to the $R_{200m} \simeq R_\mathrm{vir}$ that we use).  Knowing this, we linearly interpolate between a steepening of a factor of 1 at $R_{500}$ and a factor of 1.75 at $R_{200m}$, and integrate the resulting density profile out to each galaxy's clustercentric radius to give an updated local ICM density decreased according to \citet{morandi2015}.  The result of this is shown with the labelled line in Fig.~\ref{fig:dens_shift}.  Again, the broken-powerlaw trend is apparent and is preferred over a single power-law fit to the data by the AIC.

\subsubsection{Effect of projected radii}
\label{sec:deproject_rad}

\begin{figure}
	\centering
	\includegraphics[width=\columnwidth]{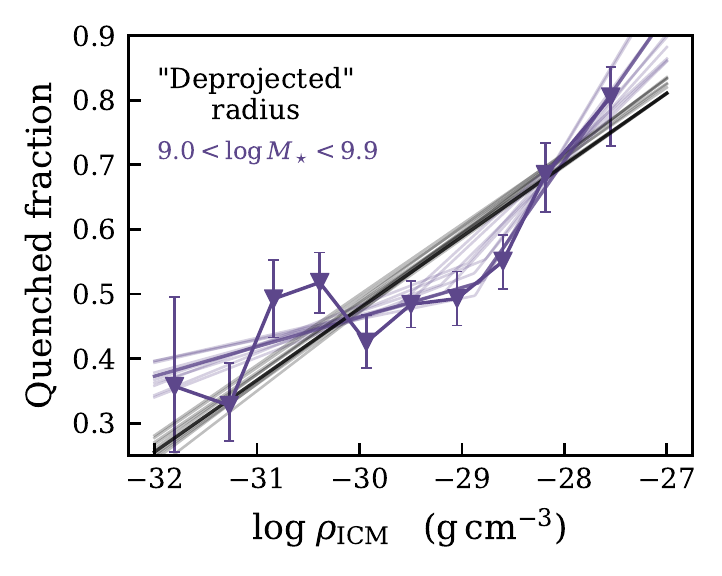}
	\caption{Quenched fraction versus ICM density where local ICM densities are determined using `deprojected' cluster-centric radii.  Best-fit single- and broken-powerlaws (gray \& black, respectively, for various rebinnings) are shown with the solid lines.}
	\label{fig:qFrac_dens_rshift}
\end{figure}

The local ICM density estimates in the paper are determined using observed projected cluster-centric radii and observed deprojected ICM density profiles.  Observing galaxy positions in projection is unavoidable, however we can attempt to gauge the effect of observing galaxy positions in projection with the aid of simulations.  In particular, observed projected radii are a lower limit to the true cluster-centric radius, and small cluster-centric radii (where ICM densities are highest) is where projection effects are most severe.  Given that strong mass segregation is not observed in low-redshift clusters \citep[e.g.][]{roberts2015, kafle2016}, projection effects should not bias the low-mass galaxies differentially relative to the higher-mass bins.  Therefore, the fact that we see trends which differ in shape between mass bins suggests that projection effects alone are not driving these differences.  However, in the interest of completeness, we explore (in an approximate manner) the influence of projected cluster-centric radii on the BPL trend seen for low-mass galaxies.
\par
To roughly deproject observed, projected cluster-centric radii we make use of dark matter simulations to obtain three-dimensional radii information.  In particular, we again make use of high-resolution dark matter only simulations from \cite{joshi2016} that were used to estimate subhalo infall velocities in section~\ref{sec:slow-then-rapid}.  We extract $R_\mathrm{3D,sim} / R_\mathrm{proj,sim}$ for galaxy-mass subhalos in these simulated galaxy clusters and then measure the median value of $R_\mathrm{3D,sim} / R_\mathrm{proj,sim}$ in bins of projected cluster-centric radius:  
\begin{equation*}
\begin{split}
R_\mathrm{proj,sim} = [0.0, 0.1, 0.2, 0.3, 0.4, 0.5, 0.6, \\
0.7, 0.8, 0.9, 1.0]\,R_{200m},
\end{split}
\end{equation*}
\noindent
for these radial bins, the median $R_\mathrm{3D,sim} / R_\mathrm{proj,sim}$ are:
\begin{equation*}
\begin{split}
\frac{R_\mathrm{3D,sim}}{R_\mathrm{proj,sim}} = [2.49, 1.55, 1.32, 1.21, 1.16, 1.11, \\
1.08, 1.07, 1.04, 1.01].
\end{split}
\end{equation*}
\noindent
When computing projected radii from the simulations we project clusters along a random axis 1000 times and then use median projected radii from these random trials.  We perform a linear spline fit to the $R_\mathrm{3D,sim} / R_\mathrm{proj,sim}$ vs. $R_\mathrm{proj,sim}$ relationship and then deproject the observed cluster-centric radius, $R_\mathrm{proj,obs}$, as follows:

\begin{equation}
    R_\mathrm{deproj,obs} = S(R_\mathrm{proj,obs} / R_{200m}) \times R_\mathrm{proj,obs}
\end{equation}
\noindent
where $S(R_\mathrm{proj,obs} / R_{200m})$ is the spline fit to the $R_\mathrm{3D,sim} / R_\mathrm{proj,sim}$ vs. $R_\mathrm{proj,sim}$ relationship interpolated to the observed normalized cluster-centric radius, $R_\mathrm{proj,obs}$ is the observed physical cluster-centric radius, and $R_\mathrm{deproj,obs}$ is the resulting estimate of the deprojected cluster-centric radius.
\par
In Fig.~\ref{fig:qFrac_dens_rshift} we show the quenched fraction as a function of ICM density, using local ICM densities determined using the deprojected cluster-centric radii outlined above.  For low-mass galaxies the BPL shape is still apparent, despite increased scatter at low-densities.  In Fig.~\ref{fig:qFrac_dens_rshift} we also show the best-fit single (gray) and broken (purple) powerlaw fits to the ``deprojected" data.  The AIC prefers the BPL fit.  We emphasize that these are rough deprojections and are only appropriate in an average sense.  Unfortunately, obtaining accurate deprojected radii on a galaxy-by-galaxy basis is not possible.  A more detailed analysis should include phase space information to incorporate deprojection not only as a function of radial position, but also as a function of velocity offsets.  We note that if we deproject only along the z-axis of the simulation, instead of many random halo projections, then the BPL trend in Fig.~\ref{fig:qFrac_dens_rshift} becomes less apparent and is only marginally preferred over a single power law fit.  It is clear that robustly deprojecting observed radii is still an outstanding issue, and we do not rule out that projection effects could contribute to the observed BPL trend.

\subsection{A mass dependence of quenching mechanisms?}

The two most commonly invoked mechanisms to quench satellite star formation in galaxy clusters are ram pressure stripping and starvation \citep{gunn1972, quilis2000, wetzel2013, muzzin2014, fillingham2015, jaffe2015, peng2015, wetzel2015}.  It is often argued that the timescale over which these two mechanisms act provides a method of disentanglement.  For instance, starvation should quench star formation on relatively long timescales ($\gtrsim 3\,\mathrm{Gyr}$ with the timescale becoming longer for low-mass galaxies) dictated by the gas depletion time, whereas efficient ram pressure stripping will deplete the galaxy of gas, and therefore quench star formation, on much shorter timescales ($\lesssim 1\,\mathrm{Gyr}$ for efficient ram pressure stripping, \citealt{quilis2000, roediger2005, steinhauser2016}).  However, this argument can be complicated by the fact that ram pressure stripping may require a significant delay before infalling galaxies encounter the densest regions of the ICM, this delay (which should be on the order of the dynamical time of the cluster) can lead to a total quenching time since infall (delay + quenching) which is similar to that of starvation.  The quenched fraction trend for low-mass galaxies in this work is consistent with gas depletion (starvation) driving the slow-quenching phase at low ICM density, and ram pressure driving the rapid-quenching phase in the cluster interior. The high-mass galaxies in this sample, though not investigated in detail in this paper, lack the same ram pressure signature.  It has been shown that gas depletion times are shortest for high-mass galaxies \citep{dave2011, fillingham2015, saintonge2017}, it is therefore plausible that high-mass galaxies consume their gas reserves and quench via starvation prior to reaching the densest cluster interior where ram pressure becomes efficient.  We also note that the deeper potential wells of high-mass galaxies will make them more resistent to ram pressure stripping in general.  Indeed, \citet{yun2018} show that the fraction of jellyfish galaxies undergoing strong ram pressure stripping in the Illustris-TNG simulation is strongly dependent on stellar mass, with the jellyfish fraction being highest for low-mass galaxies.  Furthermore, the quenching of high-mass galaxies may be largely driven by internal mechanisms, irrespective of environment \citep{peng2010}.  Star formation in low-mass galaxies should persist for much longer after cluster infall, due to the long total gas depletion times, and therefore low-mass galaxies can be still actively forming stars when reaching the densest region of the ICM; where any residual star formation may be quickly quenched due to ram pressure stripping.  Substantial ram pressure stripping of atomic gas (the more concentrated molecular component is left largely un-stripped in our models) can disconnect a galaxy from its cold-gas supply, leaving the galaxy to quench via gas depletion \citep{cen2014}.  Molecular gas depletion timescales for star-forming, low-mass galaxies are $\sim 1\,\mathrm{Gyr}$ \citep{saintonge2017}, consistent with the rapid quenching timescale we derive.  The picture that we present here is consistent with \citet{vanderburg2018} who suggest a quenching scenario where ram pressure is able to ``finish the job'' when starvation does not quench satellites rapidly enough.
\par
A mass dependent transition between starvation and ram pressure stripping has been previously advocated, where it has been argued that dwarf galaxies ($M_\star \lesssim 10^8\,\mathrm{M_\odot}$) are primarily quenched through ram pressure, whereas the quenching of galaxies with $M_\star \gtrsim 10^8\,\mathrm{M_\odot}$ is dominated by starvation \citep{fillingham2015, wetzel2015, rodriguez-wimberly2018}.  These conclusions are derived from observations of galaxies primarily in group-mass systems, which are significantly lower mass than the sample of large clusters in this work.  Here we find evidence that this transition mass may be larger ($\sim 10^9 - 10^{10}\,\mathrm{M_\odot}$) in dense clusters, where both the ICM density and relative velocities are large leading to a strong ram pressure force. 

\section{SUMMARY}
\label{sec:summary}

In this work we have used a sample of 24 low redshift SDSS galaxy clusters observed by \textit{Chandra} to present the first direct study, using a large sample of cluster galaxies, of the relationship between satellite quenching and measured ICM density.  The main results of this paper are the following:
\begin{enumerate}
  
\item Comparing quenched fractions of galaxies at the lowest ICM densities to those for isolated, field galaxies, we find evidence that approximately one third of cluster galaxies may have been pre-proccd essed prior to infall.

\item The quenched fractions of intermediate- and high-mass cluster galaxies show a modest, continuous increase with ICM density.

\item The quenched fraction vs. ICM density trend for low-mass galaxies shows evidence of a broken powerlaw trend.  The quenched fraction increases modestly at low ICM density, before increasing sharply beyond a threshold ICM density.  We show that a broken powerlaw gives a statistically better fit (even after accounting for extra parameters) than a single powerlaw.

\item The observed broken powerlaw trend is still apparent after observed cluster-centric radii are deprojected using galaxy cluster dark matter simulations, however the strength of the broken powerlaw trend shows some dependence on how galaxy positions are deprojected.  We do not rule out that projection effects may contribute to the observed trend.
  
\item The quenched fraction upturn at high ICM density, for low-mass galaxies, is well matched by a simple analytic model of ram pressure stripping, where quenching is efficient when more than $\sim$half of a galaxy's cold gas reservoir becomes susceptible to stripping.

\item These results are consistent with a slow-then-rapid picture of satellite quenching.  We argue that the slow-quenching portion is consistent with quenching via steady gas depletion (starvation) and the rapid-quenching portion is consistent with ram pressure stripping ``finishing off'' the quenching of low-mass satellites.

\end{enumerate}

\acknowledgments
IDR, LCP, JHL, and JW are supported by the National Science and Engineering Research Council of Canada.  We thank Adam Muzzin for his helpful comments on an early draft, we also thank Katy Rodriguez Wimberly for sharing the colour scheme used in figures throughout the manuscript.
\par
The scientific results reported in this article are based on data obtained from the Chandra Data Archive.  This research has also made use of software provided by the Chandra X-ray Center (CXC) in the application packages CIAO and Sherpa.
\par
Funding for the Sloan Digital Sky Survey IV has been provided by the Alfred P. Sloan Foundation, the U.S. Department of Energy Office of Science, and the Participating Institutions. SDSS-IV acknowledges support and resources from the Center for High-Performance Computing at the University of Utah. The SDSS web site is www.sdss.org.
\par
SDSS-IV is managed by the Astrophysical Research Consortium for the 
Participating Institutions of the SDSS Collaboration including the 
Brazilian Participation Group, the Carnegie Institution for Science, 
Carnegie Mellon University, the Chilean Participation Group, the French Participation Group, Harvard-Smithsonian Center for Astrophysics, 
Instituto de Astrof\'isica de Canarias, The Johns Hopkins University, 
Kavli Institute for the Physics and Mathematics of the Universe (IPMU) / 
University of Tokyo, the Korean Participation Group, Lawrence Berkeley National Laboratory, 
Leibniz Institut f\"ur Astrophysik Potsdam (AIP),  
Max-Planck-Institut f\"ur Astronomie (MPIA Heidelberg), 
Max-Planck-Institut f\"ur Astrophysik (MPA Garching), 
Max-Planck-Institut f\"ur Extraterrestrische Physik (MPE), 
National Astronomical Observatories of China, New Mexico State University, 
New York University, University of Notre Dame, 
Observat\'ario Nacional / MCTI, The Ohio State University, 
Pennsylvania State University, Shanghai Astronomical Observatory, 
United Kingdom Participation Group,
Universidad Nacional Aut\'onoma de M\'exico, University of Arizona, 
University of Colorado Boulder, University of Oxford, University of Portsmouth, 
University of Utah, University of Virginia, University of Washington, University of Wisconsin, 
Vanderbilt University, and Yale University.
%


\software{This work was made possible due to a large number of open-source software packages, including: \textsc{AstroPy} \citep{astropy2013}, \textsc{Colossus} \citep{diemer2017}, \textsc{Matplotlib} \citep{hunter2007}, \textsc{NumPy} \citep{vanderwalt2011}, \textsc{Pandas} \citep{mckinney2010}, \textsc{Photutils} \citep{bradley2016}, \textsc{SciPy} \citep{jones2001}, and \textsc{Topcat} \citep{taylor2005}}.



\appendix

\section{Ram pressure model assumptions \& considerations}
\label{sec:model_appendix}

The ram pressure model described in the manuscript makes a number of simplifying assumptions.  Below we highlight and briefly discuss the primary assumptions that went in to the ram pressure model.

\begin{itemize}[label={$-$}]

\item Both the spectral deprojection software and the fact that we fit to azimuthally averaged density profiles, assume that the clusters in the sample are spherically symmetric.  The majority of clusters in this sample do indeed show relaxed, symmetric X-ray morphologies, and if we exclude the 5/24 clusters (containing $\sim 15$ per cent of the low-mass galaxy sample) with clear signs of disturbed morphologies we find that the observed trends are unchanged.

\item We assume that the fits to the cluster density profile are valid out to the virial radius.  Previous work \citep{morandi2015} has shown that ICM density profiles for galaxy clusters tend to steepen beyond $R_{500}$, therefore we may be overestimating the local ICM density for galaxies at $R > R_{500}$.  We address this point in detail in section~\ref{sec:dens_steep}.

\item We assume that the atomic gas component is distributed in an exponential disc.  While this is a common assumption, previous work has suggested that \textsc{Hi} profiles may in fact flatten at small radii \citep{wang2014b} and flare at large radii \citep{kalberla2009, obrien2010}.  We employ an exponential profile due to the ease of constructing the analytic surface density given disc masses and scale lengths, however this is at the cost of a more realistic atomic gas distribution.

\item By assigning `pre-infall' gas masses (atomic and molecular) based on $z \approx 0$ observations, we are assuming that the gas fractions for low-mass isolated galaxies at  $z \approx 0$ are representative of typical gas fractions at the redshift of infall.  Detailed observations of total gas content for large samples of galaxies out to high redshift are not currently feasible, however semi-analytic work has found no strong evolution in the $f_\mathrm{gas} - M_\star$ relationship out to at least $z \sim 0.5-1$ \citep{popping2014} and the lookback time to these redshifts are well in excess of typical crossing times for clusters in our sample.

\item We are assuming that galaxies have not undergone significant stellar stripping since infall, consistent with simulations which show that stellar mass loss after infall is only a minor effect for cluster galaxies \citep{mistani2016, joshi2018}.

\item We are assuming that the present-day structural properties (ie.\ the outputs of the bulge+disc decompositions) of the galaxies have not evolved since infall.  Given that the fraction of bulge-dominated galaxies is enhanced in dense clusters \citep[e.g.][]{wilman2012}, the galaxies in our sample may have had smaller bulge-to-total ratios at infall.  This being the case would lead to our ram-pressure model overestimating the bulge contribution to the restoring potential of the galaxy.

\item We are assuming that galaxies interact with the ICM face-on, leading to the maximal ram pressure efficiency.  In reality, inclined interactions will reduce the efficiency of stripping, in particular for near edge-on interactions \citep{jachym2009}.  This means that our model overestimates the amount of stripping in this respect, however the edge-on interactions where this effect plays a significant role, are relatively rare.

\item We are assuming that satellite galaxies are quenched exclusively on their first infall.  While some infalling satellites will certainly survive over multiple orbits, in particular those on tangential orbits, simulations suggest that most cluster satellites are indeed quenched during first infall \citep[e.g.][]{lotz2018}.
\end{itemize}




\bibliographystyle{aasjournal}
\bibliography{manuscript.bib}



\end{document}